\documentclass[conference]{IEEEtran}
\IEEEoverridecommandlockouts

\usepackage{cite}
\usepackage{amsmath,amssymb}
\usepackage{booktabs}
\usepackage{graphicx}
\usepackage[export]{adjustbox}
\usepackage{xcolor}
\usepackage{url}
\usepackage{microtype}
\usepackage[T1]{fontenc}
\usepackage[utf8]{inputenc}
\usepackage{multirow}
\usepackage{listings}
\usepackage[hidelinks]{hyperref}
\hypersetup{
  pdftitle={Object-to-Source Mapping under Optimization: What a Control-Flow Trace Can Show, and Where Source-Level Coverage Breaks},
  pdfauthor={Alexander Weiss and Albert Schulz},
  pdfsubject={Object-to-source mapping for trace-based structural coverage under compiler optimization},
  pdfkeywords={structural coverage, MC/DC, embedded trace, object-to-source mapping, source attribution, compiler optimization, instruction trace, DO-178C}
}

\newif\ifdraftmode

\ifdraftmode
  
  \newcommand{\note}[1]{\textcolor{blue}{\small\textsf{[#1]}}}
\else
  
  \newcommand{\note}[1]{}
\fi

\begin{document}

\title{Object-to-Source Mapping under Optimization:\\
What a Control-Flow Trace Can Show,\\ and Where Source-Level Coverage Breaks}

\author{\IEEEauthorblockN{Alexander Weiss and Albert Schulz}
\IEEEauthorblockA{\textit{Accemic Technologies GmbH} \\
Kiefersfelden, Germany \\
\{aweiss, aschulz\}@accemic.com}
\thanks{Preprint, version v1, 29 August 2026. This document is a revised and
extended author version of an article first published on 17 August 2026 in
the knowledge section of the Accemic Technologies website
(\url{https://accemic.com/knowledge/object-to-source-mapping/}). The
mechanism description in Section~\ref{sec:anchor} follows the published
international patent application WO~2026/162831~A1~\cite{wo2026};
Appendix~\ref{app:asm} documents the compiler output for the worked
examples; the complete artifacts ship with the arXiv source package.}}

\maketitle

\begin{abstract}
Within a validated interval, control-flow trace can reconstruct machine-level
execution, but source-level structural coverage additionally requires a
validated attribution from generated instructions and branch outcomes to
source obligations. This paper defines that boundary and illustrates it with
exact compiler output. Optimization can fuse multiple source conditions into
one branch, replace a branch with a conditional select or move, and degrade
provenance in debug and mapping information. In such cases, trace may be
complete for the generated binary while condition-level source evidence
remains unresolved. We therefore separate trace completeness, source
attribution, and coverage satisfaction; distinguish source-to-object
transformations from object-region attribution states; and explain when
internal reconvergence makes aggregate branch observations insufficient and
ordered per-instance condition-evaluation histories necessary. We compare
measuring a less-optimized build, constraining code generation, and adding
state-writing instrumentation. Finally, we describe a pre-compilation supplementation
mechanism disclosed in WO~2026/162831~A1 that preserves selected source
distinctions as trace-observable control flow without writes to
application-resident coverage state. An appendix demonstrates the mechanism
end-to-end on real silicon for the worked example: the supplemented
\texttt{-O3} build keeps each condition individually trace-observable,
per-instance histories and an MC/DC verdict are reconstructed from a
captured ETM trace, and code-size and cycle costs are measured. Its use
remains subject to build-specific preservation, final-image mapping,
trace-validity, and probe-effect verification.
\end{abstract}


\section{Introduction}

For an observation interval whose trace-validity checks pass, control-flow
trace can provide the machine-level program-counter sequence required by the
analysis. This statement is limited to the declared trace source, protocol,
configuration and interval; detected overflow, discontinuity or unsupported
events invalidate or qualify the affected interval. Everything in this article
follows from what that sentence includes, and from what it leaves out.

\subsection{The trust chain, stated once}
\label{sec:trustchain}

A source-level result depends on a chain: source revision $\rightarrow$
compiler and flags $\rightarrow$ generated object code $\rightarrow$ linked
executable $\rightarrow$ debug and mapping information~\cite{dwarf5}
$\rightarrow$ loaded image $\rightarrow$ trace observation $\rightarrow$
reconstructed addresses $\rightarrow$ source attribution. A mismatch or
ambiguity at any link limits the source-level claim even when the object-code
observation is valid. The link stage belongs to the chain: link-time
optimization, section elimination, identical-code folding, linker relaxation
and linker-generated control flow all modify the image that is finally
observed.

Before mapping, the workflow verifies that the program image and debug
information correspond to the executable used in the run --- and that decoder,
trace configuration and mapping artifact refer to that same image. The check
uses build identifiers or cryptographic hashes rather than filenames alone.

\subsection{Terminology: condition-evaluation histories}
\label{sec:histories}

Throughout this article, a \emph{condition-evaluation history} records only
conditions that were actually evaluated. It is a sequence --- equivalently, a
vector over $\{$true, false, not evaluated$\}$ --- rather than an
always-complete Boolean vector, because short-circuit evaluation can leave
conditions unevaluated. MC/DC evaluation additionally depends on the chosen
MC/DC variant: unique-cause and masking MC/DC apply different independence
rules to the same histories~\cite{gccgcov,kvalsvik2025}. This article stays
neutral on the variant: the observation layer discussed here supplies
per-instance histories, and a downstream evaluator implements a named variant;
which variant is acceptable is settled by the applicable standard and the
agreement with the certification authority.

\section{Related work}
\label{sec:related}

Compiler-integrated, source-based coverage takes the opposite route to the
one examined here: the compiler instruments and maps at AST or preprocessor
level, as in Clang's source-based code coverage with its MC/DC
support~\cite{clangsbcc}, and GCC implements masking MC/DC on the
control-flow graph~\cite{kvalsvik2025}. These two mechanisms should not
be conflated. Clang's source-based coverage generates its source-to-counter
mapping from AST and preprocessor information before the optimizer runs, and
the inserted counter and MC/DC-bitmap updates are exactly the state-writing
instrumentation this article excludes; its documentation states that
optimizations do not affect report quality because the optimizer cannot
remove the instrumentation~\cite{clangsbcc}. GCC's condition coverage
instead derives condition outcomes from control-flow-graph
edges~\cite{kvalsvik2025}, and GCC's documentation warns that optimization
merges source constructs~\cite{gccgcov}. In our reproduction of the worked
example of Section~\ref{sec:example1}, GCC~15.2 with
\texttt{-fcondition-coverage} at \texttt{-O1} and above reports two
condition outcomes for the fused decision where \texttt{-O0} reports four
--- the fusion propagates into the compiler's own MC/DC evidence --- while
Clang~22.1 with \texttt{-fcoverage-mcdc} preserves both conditions as
separately instrumented tests, for either source form. Neither resulting
coverage report exposes the optimization-induced fusion as a distinct
diagnostic. In the tested GCC case, the report reflects the fused CFG; the
reduction from four to two condition outcomes becomes apparent when
optimization levels are compared or the generated CFG or object code is
inspected. Clang's source-based MC/DC instrumentation preserves the two
source conditions in the instrumented build, so the fusion does not arise
there. Both mechanisms are also sensitive to source representation: semantically equivalent rewrites
using a bitwise operator, a hand-folded comparison, or a conditional
expression change or eliminate the multi-condition decision recognized by
the respective coverage mechanism. In the tested cases, neither tool reports
the original two-condition MC/DC obligation; the report reflects the
rewritten source or control-flow graph rather than marking that original
obligation as uncovered. Exact sources, commands, runtime coverage outputs
(GCC~15.2.0, Clang~22.1.4) and stable Compiler Explorer links are provided
as an ancillary file (\texttt{anc/reproducibility.md}) in the arXiv source
package. On the attribution side, Stinnett
and Kell examine how far compiler-generated debugging information can carry
coverage metrics for optimized builds~\cite{stinnett2024} --- the same trust
boundary that Section~\ref{sec:hazards} discusses. Trace-based observation
itself, its protocols and its bandwidth constraints are surveyed
in~\cite{trace2022}. The present article sits between these bodies of work:
it asks what a control-flow-only trace can support without state-writing
instrumentation, where optimization breaks the passive route, and what a
pre-compilation supplementation mechanism~\cite{wo2026} changes about that
boundary. An integration-first coverage process built on embedded trace and
the same supplementation idea is described in the authors' earlier
preprint~\cite{integrationfirst}.

\section{What the protocol transmits}

The trace unit does not report every instruction. For a declared configuration
and a valid interval, it emits compressed branch-outcome and address
information --- together with synchronization points --- from which a decoder
that holds the program image reconstructs the executed control-flow sequence,
inferring the sequential instructions in between. Note that ``discontinuity''
alone would be too narrow a model: a conditional branch that is not taken
produces no program-counter discontinuity, yet its outcome is still encoded,
because the decoder needs it for the reconstruction.

The following examples use different encoding models and configuration
vocabularies; they are not interchangeable wire formats. Table~\ref{tab:protocols}
gives the logical information carried, not a complete encoding description.

\begin{table*}[!t]
\caption{What the trace protocols carry for control flow. The rows give the
logical information carried, not a complete encoding description.}
\label{tab:protocols}
\centering
\footnotesize
\begin{tabular}{@{}p{3.4cm}p{5.6cm}p{2.7cm}p{3.0cm}@{}}
\toprule
\textbf{Protocol} & \textbf{Conditional branch} & \textbf{Indirect branch} & \textbf{Resynchronization} \\
\midrule
\textbf{Arm\textregistered{} CoreSight\texttrademark{} ETMv3}~\cite{etm35} &
P-header elements carrying taken/not-taken outcomes & branch address packets &
A-Sync byte alignment plus I-Sync instruction-address and state packets \\
\addlinespace
\textbf{Arm\textregistered{} CoreSight\texttrademark{} ETMv4}~\cite{etm4} &
Atom elements; interpretation requires the reconstructed instruction context
--- do not equate E/N mechanically with taken/not-taken & address packets &
trace-synchronization and address packets \\
\addlinespace
\textbf{Intel\textregistered{} PT}~\cite{intelsdm} & TNT packets carrying
sequences of taken/not-taken outcomes & TIP packets & PSB packet-stream
boundaries with their synchronization context \\
\addlinespace
\textbf{Nexus (IEEE-ISTO 5001)}~\cite{nexus5001} & branch-history field
carrying multiple outcomes & unique address field & sync message \\
\addlinespace
\textbf{RISC-V\textregistered{} E-Trace}~\cite{etrace} & branch maps carrying
multiple branch outcomes in order; optional modes change which outcomes are
transmitted explicitly & address, differentially encoded & synchronization
packets \\
\bottomrule
\end{tabular}
\end{table*}

One compactly coded outcome per conditional branch, an address when the target
cannot be inferred --- that is the minimum control-flow idea. Actual streams
can also carry further trace information --- context, exceptions, events,
timestamps, filtering state, data trace and synchronization --- according to
architecture and configuration.

Infineon\textregistered{}'s MCDS is deliberately absent from
Table~\ref{tab:protocols}: the message-level MCDS documentation available to
the authors is subject to a non-disclosure agreement, so the wire format does
not belong in a public document, and this article makes no claim about what
it carries beyond what Infineon has published itself. The observability argument that follows is made
for control-flow-only trace configurations in general; whether and how it
applies to a specific MCDS configuration has to be established against
Infineon's documentation for that configuration.

\section{Branches are observable, conditions are not}
\label{sec:branches}

This is the statement the rest of the article rests on:

\begin{quote}
\textbf{Within a valid, continuous and correctly configured instruction-trace
interval, the decoder can reconstruct the control-flow distinctions
represented by that architecture and configuration.} Optional data and
context channels are separate claims.
\end{quote}

What a decoder can establish from the stream, with certainty:

\begin{itemize}
\item that a \textbf{branch instruction} executed;
\item \textbf{which way} it went;
\item the \textbf{target} of an indirect branch;
\item the \textbf{order} in which all of this happened.
\end{itemize}

What a control-flow-only stream does not carry:

\begin{itemize}
\item the \textbf{value} of any register, flag or memory location;
\item whether a \textbf{conditional non-branch instruction} selected its
      result --- the trace shows that the instruction executed, not what it
      decided;
\item which \textbf{source-level condition} contributed to a branch outcome;
\item the truth value of a source-level condition that did not obtain a
      \textbf{distinct, attributable, trace-observable representation} in the
      generated image.
\end{itemize}

Nothing here is a shortcoming of a particular protocol. It is the design
point: the control-flow protocols above deliberately do not carry a complete
value history of registers and memory --- continuous general data trace is
far more bandwidth- and storage-intensive than compressed branch outcomes,
which is exactly why these streams are compressed the way they
are~\cite{trace2022}.

\subsection{The same condition, twice}

\begin{figure*}[!t]
\centering
\includegraphics[scale=0.8,max width=\textwidth]{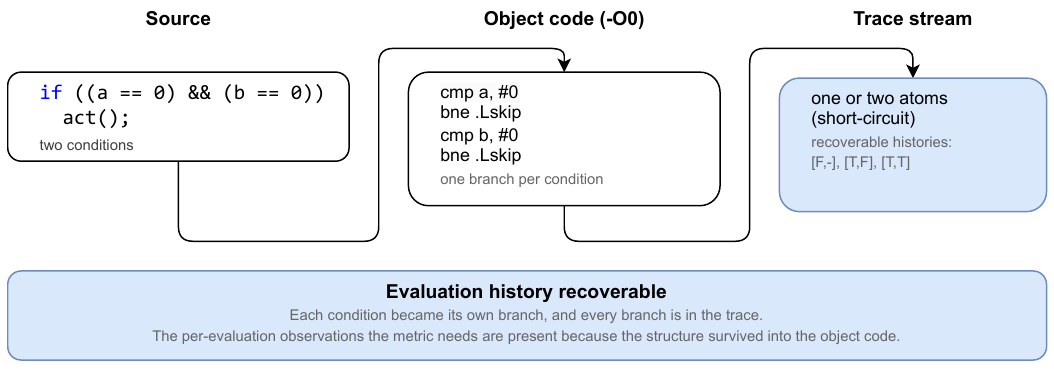}
\caption{Each executed condition is its own attributable branch, so the trace
carries one or two conditional-branch observations per evaluation, depending
on short-circuit evaluation, and the ordered evaluation history ---
$[F,-]$, $[T,F]$ or $[T,T]$ --- is recoverable. (Schematic;
Appendix~\ref{app:asm} carries the compiler output.)}
\label{fig:observable}
\end{figure*}

The two figures below (Fig.~\ref{fig:observable} and
Fig.~\ref{fig:notobservable}) are the whole argument in one picture each. Same
source line, same metric, one optimization level apart --- and the difference
is not in the trace, it is in what reached the object code.

In the first case each condition became its own conditional branch. Depending
on the outcome of the first condition, the trace contains one or two
conditional-branch observations: short-circuit evaluation ends the decision
early when the first condition already fails. When each executed condition
remains represented by a distinct attributable branch, the ordered evaluation
history --- including whether a condition was not evaluated
(Section~\ref{sec:histories}) --- is recoverable, which is the per-instance
input MC/DC~\cite{do178c} evaluation needs. In the second case the compiler
combined both conditions into one operation: the decision is still observable,
the conditions are not. Nothing was lost in transmission. The branch that
would have carried the information was never emitted. Appendix~\ref{app:asm}
carries the compiler output for this example.

\textbf{One exception worth knowing --- with narrow scope.} The ETMv4
architecture defines an optional conditional-instruction tracing
capability~\cite{etm4}. It is implementation- and instruction-set-specific:
its applicability must be established for the particular processing element,
ETM architecture revision, instruction class and trace configuration, and it
must not be assumed to expose the selected value of an A64 \texttt{csel} or an
x86 conditional move (Section~\ref{sec:example2}).

\subsection{Source attribution}
\label{sec:attribution}

Source attribution is the step between machine observation and source-level
evidence. It asks whether an observed instruction, branch outcome or path can
be assigned to the correct source-level coverage obligation without ambiguity.

This is not the same as trace completeness. An instruction trace may contain
every required machine-level event while optimization has removed the
one-to-one relationship needed to determine which source condition produced
that event.

A useful report should therefore distinguish three properties:

\begin{enumerate}
\item \textbf{Trace completeness:} whether the relevant machine execution was
      observed.
\item \textbf{Source attribution:} whether the observation can be assigned to
      the correct source obligation.
\item \textbf{Coverage achieved:} whether the attributable obligation was
      exercised as required.
\end{enumerate}

Treating these properties as one percentage can hide important limitations in
the resulting evidence. Operationally, a report can state them as separate
result fields: whether the observation interval was trace-valid (yes, no, or
qualified); how many of the total obligations are attributable
($n_A$ of $n_T$); how many attributable obligations were covered ($n_C$ of
$n_A$); how many obligations remain unresolved ($n_U$); and how many were
excluded, each with a documented justification. Unattributable obligations
are reported as unresolved; they are neither counted as covered nor silently
removed from the denominator. A measured comparison on an optimized SQLite
build --- covering instrumentation cost and source attribution --- is
published as supplementary material~\cite{kbcost}.

\section{The three questions a metric asks}

Now put the metrics next to that capability (Table~\ref{tab:metrics}). The
rows deliberately span four layers --- what the protocol reports, what the
decoder reconstructs, what a validated static mapping attributes, and what the
coverage evaluator may conclude from grouped observations --- and the
conditions column names which layers each metric needs. The condition and
MC/DC rows are answered only when \textbf{the distinctions required by the
metric have distinct, attributable, trace-observable representations in the
generated image} --- because those representations are the only place the
information can come from.

\begin{table*}[!t]
\caption{Coverage metrics against a control-flow-only stream: can control-flow
trace support the answer, and under what conditions?}
\label{tab:metrics}
\centering
\small
\begin{tabular}{@{}p{2.9cm}p{5.2cm}p{6.3cm}@{}}
\toprule
\textbf{Metric} & \textbf{The question it asks} & \textbf{Can control-flow trace support the answer, and under what conditions?} \\
\midrule
\textbf{Object instruction} & did this machine instruction execute? &
\textbf{yes} within a valid reconstructed interval --- but execution alone:
for a conditional select the chosen value, and for a predicated A32/T32
instruction even its predicate outcome, are separate questions \\
\addlinespace
\textbf{Object branch} & did both machine-level outcomes of this generated
branch occur? & \textbf{yes}, directly, within a valid reconstructed
interval \\
\addlinespace
\textbf{Source statement} & did this source-level statement execute? & only
through a validated object-to-source attribution \\
\addlinespace
\textbf{Source decision} & did both outcomes of this source-level decision
occur? & only with, in addition, a validated attribution of the object branch
to the source decision \\
\addlinespace
\textbf{Condition} & did each condition take both values? & \textbf{only if}
each condition evaluation has a distinct, attributable, trace-observable
representation --- its own branch, a supported conditional-execution event,
or a verified supplemental marker \\
\addlinespace
\textbf{MC/DC} & did each condition independently affect the outcome? &
\textbf{only if} the set of condition-evaluation histories and decision
outcomes is recoverable --- either uniquely from aggregate observations for a
non-ambiguous decision graph, or from observations grouped per execution
instance where aggregation loses correlation
(Section~\ref{sec:treediamond}) --- and the test set satisfies the
independence rule of the chosen MC/DC variant (Section~\ref{sec:histories}) \\
\bottomrule
\end{tabular}
\end{table*}

That is the observability problem in one sentence: \textbf{the relevant
distinctions have to be trace-observably represented in the generated image.}
Whether they are is decided by the compiler, not by the tool.

\section{How mapping results are classified}
\label{sec:classes}

Mapping results are best described along two separate axes, because a single
class list mixes categories of different kinds. The first axis is the
\textbf{source-obligation transformation}: an obligation can be preserved
one-to-one, split one-to-many, merged many-to-one, cloned or inlined into
multiple contexts, or eliminated. The second axis is the
\textbf{object-region attribution status}: a region of generated code can be
directly attributable, ambiguously attributable, compiler-generated, or
unmapped. Every source obligation then carries a transformation state, and
every object-code region carries an attribution status --- without an
artificial precedence rule between overlapping class names.

\begin{figure*}[!t]
\centering
\includegraphics[scale=0.8,max width=\textwidth]{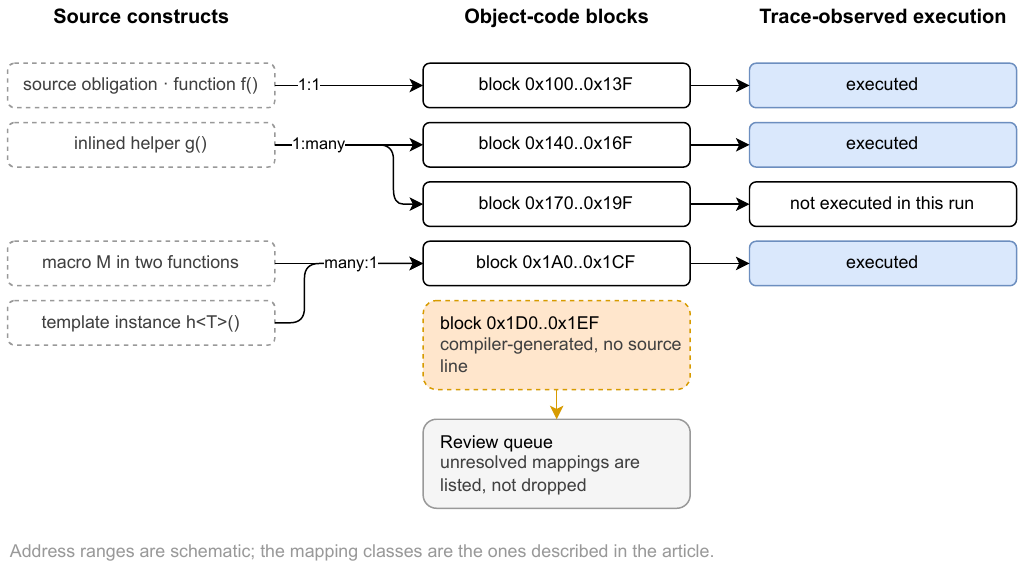}
\caption{Reporting policy: every object-code region is assigned an attribution
status and every source obligation a transformation state --- mapped
one-to-one, split, merged, or handed to the review queue --- rather than being
silently dropped (schematic).}
\label{fig:accountability}
\end{figure*}

The reporting policy assumed in this article is that every object-code region
is assigned an attribution status and every obligation a transformation state,
both surface in the report rather than being silently dropped, and source
coverage is computed only according to the documented treatment of those
states (Fig.~\ref{fig:accountability}).

\section{Example 1: the computed decision}
\label{sec:example1}

\noindent\begin{minipage}{\columnwidth}
\begin{lstlisting}
if ((a == 0) && (b == 0))
    act();
\end{lstlisting}
\end{minipage}

\textbf{At \texttt{-O0}} the compiler emits the short-circuit structure
literally: compare \texttt{a}, branch; compare \texttt{b}, branch. Two
conditions, two conditional branches (Appendix~\ref{app:asm}, Listing~B.1).

Depending on the outcome of the first condition, the trace then carries one
or two conditional-branch observations: $[F,-]$ when the first condition
already fails, $[T,F]$ or $[T,T]$ otherwise. Each executed condition is
attributable to its own branch, so the ordered evaluation history ---
including that a condition was not evaluated --- is recoverable
(Section~\ref{sec:histories}).

\textbf{At \texttt{-O3}} the compiler is free to evaluate the whole decision
arithmetically --- combine both comparisons into one operation and emit a
single conditional branch on the result. For the source above, clang~22.1.4
for AArch64 emits exactly that: \texttt{orr~w8,~w1,~w0} followed by one
\texttt{cbz} (Appendix~\ref{app:asm}, Listing~B.2).

Now the stream carries \textbf{one} conditional-branch observation per
evaluation. The decoder sees that the decision was taken or not taken. It
cannot see which condition made it so, because the information never entered
the control flow: \emph{the conditional branch is visible in the trace, the
arguments of the arithmetic operation are not.}

\begin{figure*}[!t]
\centering
\includegraphics[scale=0.8,max width=\textwidth]{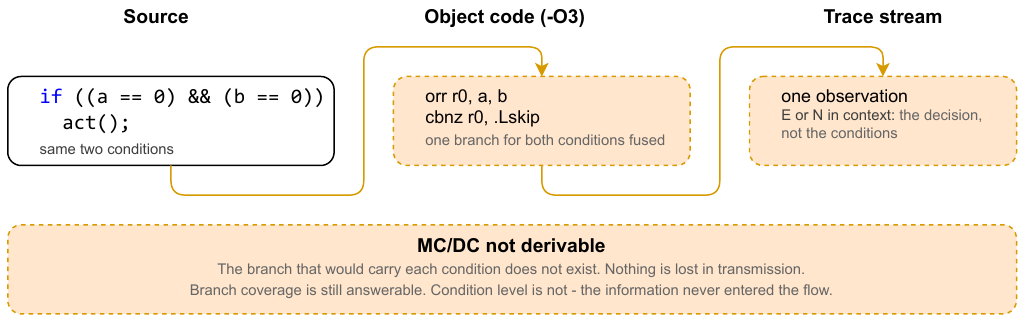}
\caption{Same source, one optimization level higher. Nothing was lost in
transmission --- the branches that would carry the individual conditions were
never emitted. (Schematic; exact output in Appendix~\ref{app:asm}.)}
\label{fig:notobservable}
\end{figure*}

Read that against Table~\ref{tab:metrics} and the verdict is not a matter of
tool quality. Object-branch coverage is still answerable. Condition coverage
and MC/DC are not, because the branches they would need do not exist
(Fig.~\ref{fig:notobservable}).

\section{Example 2: the conditional non-branch instruction}
\label{sec:example2}

\noindent\begin{minipage}{\columnwidth}
\begin{lstlisting}
b = (a == 0) ? 42 : b;
\end{lstlisting}
\end{minipage}

\textbf{At \texttt{-O0}}, again the literal structure: a compare and a
conditional branch around the assignment (Appendix~\ref{app:asm},
Listing~B.3).

\textbf{At \texttt{-O3}}, the compiler emits a conditional selection instead
of a branch: on x86-64 a \texttt{cmovne}, on AArch64 a \texttt{csel}
conditional-select instruction (Appendix~\ref{app:asm}, Listings~B.4
and~B.5). There is no branch at all. Three object-code forms have to be kept
apart here, because their observability differs: A32/T32 predicated execution
and IT blocks; A64 conditional-select instructions, which always execute and
choose an operand from the condition flags; and x86 conditional moves.

\begin{table}[!t]
\caption{Branchless object-code forms and what a control-flow-only trace
shows.}
\label{tab:branchless}
\centering
\footnotesize
\begin{tabular}{@{}p{2.2cm}p{2.6cm}p{3.0cm}@{}}
\toprule
\textbf{Object-code form} & \textbf{Control-flow-only trace} & \textbf{ETMv4 conditional-instruction option~\cite{etm4}} \\
\midrule
A32/T32 predicated instruction & pass/fail of the predicate can remain
open & possibly supported --- establish via PE documentation and
identification registers \\
\addlinespace
A64 \texttt{csel} & instruction visible, selection not & not to be assumed
--- exact ETM instruction classification required \\
\addlinespace
x86 \texttt{cmov} & instruction visible, selection not & not applicable \\
\bottomrule
\end{tabular}
\end{table}

This case is subtler than the first because execution of the instruction is
observable while the selected outcome may not be (Table~\ref{tab:branchless}).
Many common instruction-trace
configurations report that the instruction \textbf{executed}; they do not
report the \textbf{predicate outcome} --- whether the destination received the
new value or kept its old one. So the stream does not merely lack the
condition --- it contains a positive signal that can be misread as coverage.
The instruction ran. What it selected is not in the trace.

The optional ETMv4 conditional-instruction tracing
capability~\cite{etm4} is implementation- and instruction-set-specific. Its
applicability must be established for the particular processing element, ETM
architecture revision, instruction class and configuration --- and it must
not be assumed to expose the selected value of an A64 \texttt{csel} or an x86
conditional move. Where it does not apply, one directly verifiable mitigation
is the build constraint below.

The documented mitigation is a build constraint: forbid conditional non-branch
instructions. The available controls are compiler- and target-specific aids
rather than a general guarantee --- coding style helps, but avoiding the
ternary operator does not prevent a compiler from if-converting an ordinary
\texttt{if}/\texttt{else}, and flags such as GCC's
\texttt{-fno-if-conversion} and \texttt{-fno-if-conversion2}~\cite{gccopt}
address specific passes while back-end instruction selection can still produce
branchless forms. What generalizes is the verification: prove the absence of
conditional non-branch instructions in the generated object code.

Note what just happened, though. The coverage requirement has started to
dictate the code generation of the build you ship.

\section{Two further mapping hazards}
\label{sec:hazards}

\textbf{Context.} After optimizing transformations, the mapping from a source
condition to a specific conditional branch can simply be gone --- the branch
exists, its provenance does not. The blunt mitigation recorded in our own
coverage notes --- \emph{if the compiler does not optimize the code
(\texttt{-O0}), the described effect does not occur} --- is an observation
scoped to the compilers and source patterns it was recorded for, not a general
rule: \texttt{-O0} disables most discretionary optimization passes, not every
transformation~\cite{gccopt}; mandatory lowering, instruction selection and
macro expansion remain. The preprocessor makes provenance worse: common
compilers attribute every token of a multi-line macro expansion to the
invocation location, in which case column information cannot separate the
expanded tokens --- DWARF~5 defines dedicated macro-information
structures~\cite{dwarf5}, but line tables that point at the invocation site
are what tools frequently encounter. Appendix~\ref{app:asm}, Listing~B.6,
demonstrates both effects with exact compiler output: in a two-condition
macro expansion compiled at \texttt{-g -O2}, the first condition carries the
line and column of the invocation site, while the second condition ---
together with the conditional increment the optimizer formed from the inner
branch --- carries \emph{line~0}, the DWARF convention for ``no source line
attributable''. Neither line-table entry separates the two conditions. How
far compiler-generated debugging information can carry coverage claims for
optimized builds is examined in~\cite{stinnett2024}; GCC documents the
corresponding \texttt{gcov} limitations under optimization
itself~\cite{gccgcov}.

\textbf{Direction.} The compiler chooses whether a source-level \texttt{true}
becomes a jump taken or not taken --- \texttt{JZ} or \texttt{JNZ} is its
decision. This is not an information loss but a polarity-mapping task: the
source-to-object polarity must be established before a missing branch outcome
may be reported as a missing source outcome. Once both outcomes of the
object branch have been observed, polarity no
longer affects the binary branch-coverage verdict; it remains part of the
source attribution needed for diagnostics and for interpreting
condition-level histories --- and it still costs the engineer time when
hunting a missing test case.

\section{Ordered histories and internal reconvergence}
\label{sec:treediamond}

\begin{figure*}[!t]
\centering
\includegraphics[scale=0.8,max width=\textwidth]{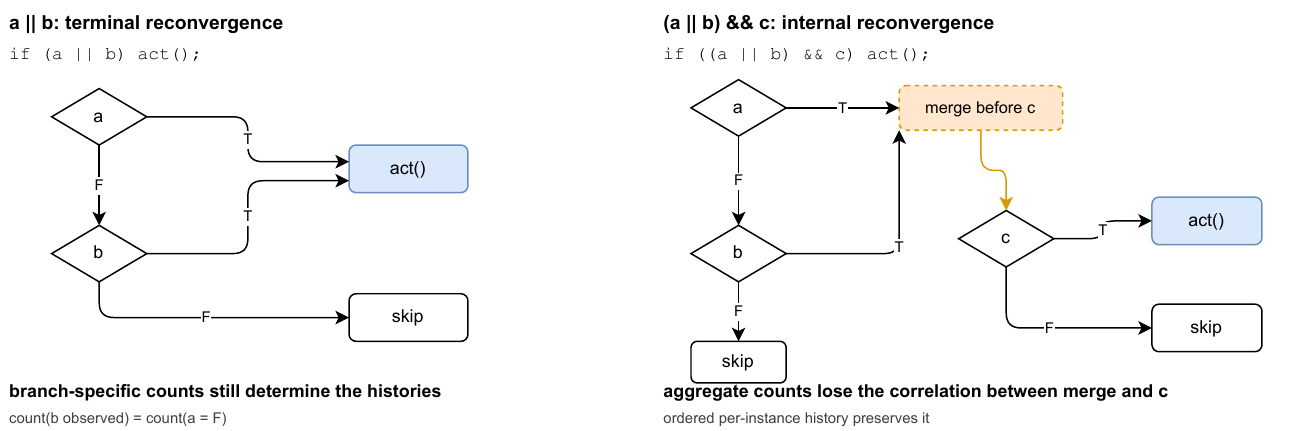}
\caption{Terminal versus internal reconvergence. Left: for \texttt{a || b},
branch-specific aggregate counts still determine the evaluation histories ---
only the final successor is ambiguous. Right: for \texttt{(a || b) \&\& c},
partial histories merge before \texttt{c} is evaluated; aggregate counts lose
the correlation between the merge and the outcome of \texttt{c}, while an
ordered per-instance history preserves it.}
\label{fig:treediamond}
\end{figure*}

Reconvergence at a final decision successor does not by itself make aggregate
branch observations insufficient. For a simple decision such as
\texttt{a~||~b}, counts for the distinct object branches still determine how
often the histories $[T,-]$, $[F,T]$ and $[F,F]$ occurred: \texttt{b} is only
evaluated when \texttt{a} is false. The final successor alone is ambiguous,
but the branch-specific aggregate observations are not
(Fig.~\ref{fig:treediamond}, left).

Ambiguity arises when different partial histories reconverge and execution
subsequently reaches another condition or split within the same source
decision. For \texttt{(a~||~b)~\&\&~c}, the histories $[T,-,*]$ and
$[F,T,*]$ merge before the evaluation of \texttt{c}. Aggregate outcome counts
for \texttt{a}, \texttt{b} and \texttt{c} do not preserve which outcome of
\texttt{c} belonged to which incoming history: the test set
$\{[T,-,T],\,[F,T,F]\}$ and the test set $\{[T,-,F],\,[F,T,T]\}$ produce
identical aggregate branch counts, yet contain different evaluation histories
(Fig.~\ref{fig:treediamond}, right).

A complete, ordered trace does preserve the observations emitted before and
after the merge. If the observations are attributable and delimited per
decision instance, the complete condition-evaluation histories remain
reconstructible. Per-instance grouping is required where the decision graph
does not permit the histories to be derived uniquely from aggregate
observations. Section~\ref{sec:anchor-evidence} describes a grouping model
that supplies exactly this delimitation.

Optimization does not make source mapping uniformly impossible. It makes the
relationship non-uniform: source statements can be reordered, merged,
duplicated by inlining or represented without a distinct address range. The
report preserves those distinctions rather than converting every executed
address into a simple line hit.

\section{Common responses}
\label{sec:stateoftheart}

Every tool that derives source-level coverage from a control-flow trace plus
debug information sits inside the boundary above. This is not a claim about
implementation quality: it is an upper bound on the information available in
the binary, and it therefore applies to every tool of the class --- ours
included.

Four response classes are common (Table~\ref{tab:escapes}); they are response
classes, not an exhaustive taxonomy --- selective data trace, source
transformation, binary rewriting and compiler-metadata approaches exist, and
whether any of them satisfies a given assurance case is a separate argument.
The fourth class is the subject of Section~\ref{sec:anchor}.

\begin{table*}[!t]
\caption{Common responses, what each can provide, and what each costs or
leaves open.}
\label{tab:escapes}
\centering
\small
\begin{tabular}{@{}p{6.1cm}p{4.0cm}p{4.7cm}@{}}
\toprule
\textbf{Response} & \textbf{What it can provide} & \textbf{Principal cost or residual obligation} \\
\midrule
\textbf{Measure a less-optimized build} & may preserve source-level decision
structure & measurement/production mismatch, or production performance and
footprint cost \\
\addlinespace
\textbf{Constrain code generation} & can suppress selected branchless
transformations & compiler- and target-specific; the final linked image must
be verified \\
\addlinespace
\textbf{Add state-writing coverage instrumentation} & direct condition and
decision observations & runtime state, writes, code/data footprint, timing
and cache effects \\
\addlinespace
\textbf{Add control-flow supplementation without application-resident
analysis state} (Section~\ref{sec:anchor}) & trace-observable distinctions in
the optimized build & code/layout/timing effects; preservation and mapping
must be verified per build \\
\bottomrule
\end{tabular}
\end{table*}

Several of these responses are not tool-independent: constraints live in the
compiler, instrumentation and supplementation in the coverage toolchain, so
the constraint and the evidence about it can come from the same place --- a
dependence a reviewer will want addressed.

\begin{quote}
\textbf{Do not infer through an absent distinction.} If the generated object
control flow does not distinguish two source outcomes and no verified
supplemental observation exists, the source-level obligation remains
unresolved.
\end{quote}

\section{A pre-compilation supplementation approach}
\label{sec:anchor}

The gap is now precisely located, which makes the approach statable in one
sentence:

\begin{quote}
\textbf{If the missing information is not in the control flow, put it there
--- as control flow, and without writes to application-resident analysis
state.}
\end{quote}

At the points a coverage metric needs to tell apart, the source is
supplemented with control-flow blocks that are

\begin{itemize}
\item \textbf{constructed to be functionally neutral} --- semantic
      preservation is verified for the named source pattern, toolchain and
      build;
\item \textbf{preserved through optimization} --- established at build time,
      per compiler, version and optimization setting
      (Section~\ref{sec:anchor-preservation});
\item \textbf{free of memory operations on analysis state} --- the
      application maintains no coverage counters and no software trace buffer
      (Section~\ref{sec:anchor-ontarget}).
\end{itemize}

This section describes the mechanism --- marketed as CEDARtools.Anchor --- as
the published international patent application WO~2026/162831~A1 discloses
it~\cite{wo2026}.

\subsection{Selecting the distinctions to preserve}
\label{sec:anchor-selection}

The supplementation step starts from an analysis objective and identifies the
source locations or expressions whose executions must remain distinguishable
for that objective. Each selected point of interest (POI) receives a unique
identifier representing a trace-observable marker or trace-correlatable anchor
point. The source-level supplementation may use non-executable or functionally
neutral constructs --- labels, neutral blocks, or dummy calls --- at those
POIs, and the POI selection may be derived from the analysis objective by
static analysis of the source code~\cite[description pp.~12, 44]{wo2026}.
Supplementation is a scoped selection rule, not blanket instrumentation of the
source base.

\subsection{Preservation through compilation and linking}
\label{sec:anchor-preservation}

The supplemented source is provided to the compiler, which generates
supplemented machine code. The optimizer may restructure, relocate or remove
ordinary source-code regions, while the protected POI sections are constructed
so that the compiler refrains from relocating or removing them from the
machine-level control flow~\cite[description pp.~12, 19]{wo2026}. The build
produces machine code with observable POI-related control flow together with
\textbf{static observation-mapping data}: a static correspondence between
source-level POIs and their binary representations, used to correlate runtime
trace events with source constructs~\cite[description pp.~44--45]{wo2026}.
This is the causal step that a downstream trace-path device cannot supply by
itself: optimization-resistant source/object distinctions exist in the binary
because they were placed and protected before and during compilation, not
retrofitted afterwards. Preservation and mapping are then verified against the
final linked and loaded image, including the effects of link-time
optimization, section elimination, code folding, linker relaxation and
linker-generated control flow (Section~\ref{sec:trustchain}).

\subsection{What remains on the target}
\label{sec:anchor-ontarget}

The POI sections avoid loads and stores to coverage counters, profiling
variables, trace-log buffers and other application-resident runtime-analysis
state~\cite[description p.~15]{wo2026}. The target provides trace messages or
an executed-instruction-address stream, while a runtime-analysis processing
unit outside the target CPU reconstructs the control flow and determines POI
execution instances at machine-code level~\cite[description pp.~19, 44]{wo2026}.
The associated mapping metadata may consist of debugging data, symbol-table
associations, or both~\cite[description p.~20]{wo2026}; the machine-level POI
results are then transformed into a source-level representation.

This is the property that separates \emph{counterless control-flow
supplementation} from \emph{state-writing instrumentation}. Supplementing the
source for observability is instrumentation in the broad sense --- the program
representation is modified for measurement. What the disclosed variant does
not do is write coverage state from the application: conventional counter- or
tag-based instrumentation performs loads and stores to analysis state, and
those accesses can touch caches, timing and the memory budget depending on
configuration. An anchor section produces no such access; it still executes as
part of the program, and its instruction-level effect is measured separately.

\subsection{From POI events to coverage evidence}
\label{sec:anchor-evidence}

For each multi-condition decision subject to MC/DC, the supplementation can
define a POI group containing POIs for the atomic condition-evaluation points
and, optionally, the decision outcome or an exit boundary. A lookup table maps
executed-instruction addresses to the group and its members, so that each
execution instance can be represented by the POIs observed during that
instance~\cite[description p.~37]{wo2026}. Where the native attributable
branch observations uniquely determine the set of evaluation histories, they
may be sufficient without additional per-instance
grouping~\cite[description p.~39]{wo2026}; where optimization removes source
distinctions, or where internal reconvergence makes aggregate observations
non-unique, the analysis records the ordered condition-evaluation history of
each instance and relates it to the resulting branch
behavior~\cite[description pp.~36, 39]{wo2026}. MC/DC is evaluated outside the
target by comparing the observed histories under the independence rule of the
chosen MC/DC variant (Section~\ref{sec:histories})~\cite[description
p.~38]{wo2026}. This is the grouping model that
Section~\ref{sec:treediamond} requires: aggregate branch counts may be
insufficient where internal reconvergence makes the set of evaluation
histories non-unique, while per-instance histories remain reconstructible
when the required POIs and their grouping exist --- the approach can make the
required distinctions recoverable.

\subsection{Build involvement, artifact identity, probe effect}
\label{sec:anchor-build}

The mechanism is applied before compilation: augmented source and its header
or macro definitions are compiled and linked into the machine code used for
analysis~\cite[description p.~45]{wo2026}. The added constructs can impose
localized timing and memory-footprint overhead and can require optimization
barriers around the selected points~\cite[description p.~13]{wo2026}.

What that changes, against Table~\ref{tab:escapes}: the optimization level
can stay where the product needs it; the measured artifact can remain the
shipped artifact --- the publication states that the impact can be small
enough to justify leaving the modified code in the
release~\cite[description p.~16]{wo2026}, in which case artifact identity
means releasing the supplemented image that was measured, under the trust
chain of Section~\ref{sec:trustchain}, or making a separate
representativeness argument.

\textbf{Probe effect.} The approach is designed to reduce the probe effect
relative to state-writing instrumentation; it is not probe-effect-free. The
supplemented machine-code structure exists, occupies code space, and executes
--- code size, control-flow and layout changes, cache effects and localized
optimization barriers are possible effects, and the timing effect must be
measured for the named target, compiler, placement and configuration. What
the supplementation avoids is reads and writes of application-resident
analysis state --- where much of the cost of conventional instrumentation
typically lives~\cite[description pp.~13--15]{wo2026}.

\textbf{Limitations.} This section describes the mechanism disclosed in
WO~2026/162831~A1. Appendix~\ref{app:anchor-demo} demonstrates the chain
end-to-end for the worked example on one named Cortex-M7 configuration and
reports cycle cost for that experiment; a separate Cortex-M4 build provides
the code-size micro-benchmark. Beyond those named configurations, the present
article does not establish tool qualification, semantic equivalence for
every source pattern, preservation under every compiler and linker
pipeline, or a universal overhead bound. Those properties require
build-specific verification and empirical evidence for the target
configuration.

\textbf{Status.} The mechanism is disclosed in the published international
patent application WO~2026/162831~A1~\cite{wo2026}.

\section{What this article establishes}

\begin{itemize}
\item trace completeness, source attribution and coverage achieved are three
      different questions, and a report can state them as separate fields
      (Section~\ref{sec:attribution});
\item source-obligation transformations and object-region attribution states
      are two separate axes, under a reporting policy in which nothing is
      silently dropped (Section~\ref{sec:classes});
\item two object-code mechanisms by which source distinctions disappear,
      together with attribution hazards caused by optimized debug and
      provenance information --- all shown with exact compiler output
      (Appendix~\ref{app:asm}, Section~\ref{sec:hazards});
\item decisions whose internal reconvergence makes aggregate observations
      non-unique require ordered per-instance condition-evaluation histories
      (Section~\ref{sec:treediamond});
\item a pre-compilation supplementation mechanism disclosed in
      WO~2026/162831~A1 can make those distinctions trace-observable without
      writes to application-resident analysis state, subject to the
      verification obligations stated in Section~\ref{sec:anchor} --- and is
      demonstrated end-to-end on real silicon for the worked example, with
      measured costs (Appendix~\ref{app:anchor-demo}).
\end{itemize}

\appendices
\lstset{basicstyle=\ttfamily\scriptsize, frame=tb, framerule=0.3pt,
  xleftmargin=0pt, framexleftmargin=0pt, aboveskip=5pt, belowskip=5pt,
  columns=fullflexible, keepspaces=true, breaklines=true}


\section{End-to-end demonstration of the supplementation mechanism}
\label{app:anchor-demo}

This appendix demonstrates the mechanism of Section~\ref{sec:anchor} on the
worked example of Section~\ref{sec:example1}. It comprises two experiments
on the same subject translation unit. The \textbf{Cortex-M7 experiment}
(NUCLEO-H7S3L8, 600\,MHz, code executing from flash) demonstrates the
trace-to-evidence chain end-to-end --- captured ETM trace, reconstructed
per-instance histories, MC/DC verdict --- and provides the cycle
measurements. A separate \textbf{Cortex-M4 build} provides the code-size
micro-benchmark and the listings in Sections~A.1--A.4; it is not part of
that trace artifact chain. Toolchain for all builds: Arm GNU Toolchain
14.2.Rel1 (\texttt{arm-none-eabi-gcc} 14.2.1), \texttt{-O3 -g3} plus the
anchor-distinctness flags \texttt{-fno-ipa-icf -fno-crossjumping
-fno-tree-tail-merge} --- applied identically to the reference and the
supplemented build, a concrete instance of the per-build verification
obligation of Section~\ref{sec:anchor-build}. The driver exercises a
minimal three-vector test set (a \emph{masking-MC/DC} set,
Section~\ref{sec:histories}) for $c_1 = (a{==}0)$, $c_2 = (b{==}0)$:
$(a,b) = (0,0) \rightarrow [T,T]$, $(0,1) \rightarrow [T,F]$,
$(1,7) \rightarrow [F,-]$.

\subsection*{A.1 Reference build without supplementation}

At \texttt{-O3} the compiler fuses both conditions into one branch,
$(a{==}0)\,\&\&\,(b{==}0) \rightarrow (a\,|\,b){==}0$ --- the Thumb-2
counterpart of Listing~B.2:

\begin{lstlisting}[language={}]
00000040 <decision>:
  40:  orrs   r0, r1
  42:  bne.n  48
  44:  b.w    act
  48:  bx     lr
\end{lstlisting}

The two source conditions are not individually observable in the control
flow.

\subsection*{A.2 Supplemented source}

The supplementation step (libclang-based, per
Section~\ref{sec:anchor-selection}) wraps each condition and each outcome
block in functionally neutral anchor constructs --- two condition POIs, two
block POIs, one MC/DC group:

\begin{lstlisting}
if (CFOP_COND(decision_L2_001, ((a == 0))) &&
    CFOP_COND(decision_L2_002, ((b == 0))))
    { cfop_pin_decision_L1_001:
      CFOP_KEEP_BLOCK_POS(decision_L1_001); act(); }
else
    { cfop_pin_decision_L1_002:
      CFOP_KEEP_BLOCK(decision_L1_002); }
\end{lstlisting}

The four source-level POIs expand into six distinct binary observation
sites: a true and a false site for each of the two conditions, plus the two
decision-outcome sites.

\subsection*{A.3 Final linked image at \texttt{-O3} (Cortex-M4 build)}

In the linked \texttt{-O3} image the fusion is decomposed into per-condition
control flow; all six observation sites are named symbols at distinct
addresses, and short-circuit evaluation is preserved --- the $c_2$
sites are reachable only after $c_1 = T$:

\begin{lstlisting}[language={}]
00000040 <decision>:
  40:  cbnz  r0, 4c
00000042 <keep_cond_..._L2_001_T>:
  42:  nop
  44:  cbnz  r1, 52
00000046 <keep_cond_..._L2_002_T>:
  46:  nop
00000048 <keep_block_..._L1_001>:
  48:  b.w   act
0000004c <keep_cond_..._L2_001_F>:
  4c:  nop
0000004e <keep_block_..._L1_002>:
  4e:  nop
  50:  bx    lr
00000052 <keep_cond_..._L2_002_F>:
  52:  nop
  54:  b.n   4e
\end{lstlisting}

(Symbol names shortened for print; the full names carry the decision id and
a site hash.)

\subsection*{A.4 Static observation mapping (Cortex-M4 build)}

The build derives the static observation mapping of
Section~\ref{sec:anchor-preservation}: per condition POI the
true/false site addresses ($c_1$: \texttt{0x42}/\texttt{0x4c}; $c_2$:
\texttt{0x46}/\texttt{0x52}), the outcome-site addresses (T:
\texttt{0x48}, F: \texttt{0x4e}), and the statically derived truth table of
observation masks, in which the mask ``$c_2$ observed without $c_1$'' is
correctly marked unreachable (short-circuit). The Cortex-M7 image realizes
the same supplementation of the same translation unit; its site addresses
appear in Section~A.5.

\subsection*{A.5 Captured trace and reconstructed histories (Cortex-M7 experiment)}

The ETM trace was routed to the CoreSight TMC configured in ETF
circular-buffer mode and read out through SWD using the board's on-board
debug adapter; no external trace port or high-bandwidth trace probe was
used. The buffer capacity is 2\,KB; the 128-byte capture of a four-sweep
run did not wrap (write pointer \texttt{0x80} within the 2\,KB buffer,
buffer size and circular mode confirmed from the read-back status
registers), and the decoded interval started from a valid synchronization
point (one contiguous segment; 43 packets, 80 decoded elements, no overflow
or discard indication). The control-flow reconstruction was cross-checked
against an independently implemented decoder oracle; both produce identical
reconstructed instruction-address sequences. The per-condition sites appear
individually --- $c_1$ taken/not-taken $4/8$, $c_2$ $4/4$: exactly four
sweeps of the three vectors, with the short-circuit visible in the
measurement ($c_2$ reached $8$ of $12$ times). The reconstructed
per-instance condition-evaluation histories, four times each:

\begin{lstlisting}[language={}]
[T,T] -> T   c1:T@0x8000b66
             c2:T@0x8000b6a  out:T@0x8000b6c
[T,F] -> F   c1:T@0x8000b66
             c2:F@0x8000b76  out:F@0x8000b72
[F,-] -> F   c1:F@0x8000b70  out:F@0x8000b72
\end{lstlisting}

The demonstrated variant is \textbf{masking MC/DC}: the independence
witness for $c_1$ is $[T,T] \rightarrow T$ versus $[F,-] \rightarrow F$,
with $c_2$ masked (not evaluated) in the latter evaluation; the witness for
$c_2$ is $[T,T] \rightarrow T$ versus $[T,F] \rightarrow F$. Across the
four sweeps, the aggregated condition-outcome counts were $c_1$: $T{=}8$,
$F{=}4$ and $c_2$: $T{=}4$, $F{=}4$; the verdict reports both conditions
covered. A deterministic second track --- the same linked binary executed
in an instruction-set interpreter whose ETMv4 emission feeds the same
decoder --- reproduces the identical histories; repeated interpreter runs
regenerate the same synthetic ETMv4 byte stream byte-for-byte.

\subsection*{A.6 Measured cost}

All figures are measured on the named builds; they are micro-benchmark
values for this one decision and carry no general overhead claim
(Table~\ref{tab:demo-cost}).

\begin{table}[!t]
\caption{Measured micro-benchmark cost of the supplementation (reference
versus supplemented build, identical distinctness flags).}
\label{tab:demo-cost}
\centering
\footnotesize
\begin{tabular}{@{}lrrr@{}}
\toprule
\textbf{Metric} & \textbf{Ref.} & \textbf{Suppl.} & $\Delta$ \\
\midrule
\texttt{.text} size, bytes (M4 build) & 336 & 348 & $+12$ \\
\quad thereof \texttt{decision}, bytes & 10 & 22 & $+12$ \\
Executed instr., 3-vector run (M4, interp.) & 36 & 43 & $+7$ \\
Cycles per 3-instance sweep (M7, min.) & 442 & 508 & $+66$ \\
\bottomrule
\end{tabular}
\end{table}

The $+12$ bytes (five anchor \texttt{nop}s of 10 bytes plus one 2-byte join
branch) are the incremental cost of supplementation relative to a reference
build using the identical distinctness flags; they do not include potential
whole-program effects of introducing those flags in a larger program. The
cycle figures (Cortex-M7 experiment; DWT cycle counter read over SWD, no
trace-port capture involved; minimum observed per sweep) correspond to an
arithmetic average of $+22$ cycles per invocation for this path mix; no
per-path overhead is inferred. Two independent runs with re-flash produced
identical values; the measuring script and raw per-run readings are
archived with the evidence package. Not measured, and stated as such:
per-path cycle attribution, cache and interrupt configuration effects,
energy, and costs on a real-scale workload. The underlying artifacts ---
reference and supplemented ELF, transformed source, mapping, raw ETM
capture, decoder outputs and measurement scripts --- are archived as a
versioned evidence package and can be shared on request.

\section{Compiler-output excerpts and reproducibility artifacts}
\label{app:asm}

The listings below reproduce the complete instruction and label sequences
relevant to the worked examples, produced by \textbf{clang version 22.1.4}
on the exact sources shown. Whitespace is normalized for print, and
compiler-generated comments and directives without semantic relevance to
the analysis are omitted. The complete, unedited \texttt{-S} outputs are
included as files in the arXiv source package. The figures in the body are
schematics of the same mechanisms. Every listing was produced with the
command
\begin{center}
\texttt{clang -target <target> -O<level>}\\
\texttt{-fno-asynchronous-unwind-tables -S <source>}\\[2pt]
\texttt{<target>} $\in$ \{\texttt{aarch64-none-elf},
\texttt{x86\_64-unknown-linux-gnu}\}
\end{center}
with the optimization level named per listing. The flag
\texttt{-fno-asynchronous-unwind-tables} is passed only to suppress unwind
directives; it does not change the generated instructions of these
functions. Listing~B.6 additionally passes \texttt{-g} (its subject is the
emitted line table) and \texttt{-fdebug-compilation-dir=.}, which only
neutralizes the embedded compilation-directory path.

Listings B.1--B.5 were additionally cross-checked on Compiler Explorer
(\url{https://godbolt.org/}, accessed 29~August 2026): armv8-a clang~22.1.0
reproduces Listings~B.1--B.4 instruction for instruction, x86-64 clang~20.1.0
reproduces Listing~B.5, and 32-bit Arm GCC~14.2\footnote{Compiler Explorer id
\texttt{carmug1420}, ARM GCC 14.2.0 (unknown-eabi), \texttt{-O3 -marm}.}
confirms the fused shape shown schematically in Fig.~\ref{fig:notobservable}
(\texttt{orrs r0, r0, r1} followed by a single conditional instruction).
Stable links preloading source, compilers and flags:
Example~1: \url{https://godbolt.org/z/q41eYqK5E} ---
Example~2: \url{https://godbolt.org/z/PKfj5M6b8}.

\subsection*{B.0 Sources}

\noindent\texttt{ex1-computed-decision.c} (Example~1):
\begin{lstlisting}
void act(void);

void f(int a, int b)
{
    if ((a == 0) && (b == 0))
        act();
}
\end{lstlisting}

\noindent\texttt{ex2-conditional-move.c} (Example~2):
\begin{lstlisting}
int g(int a, int b)
{
    b = (a == 0) ? 42 : b;
    return b;
}
\end{lstlisting}

\subsection*{B.1 Example 1 at \texttt{-O0}, AArch64}

\noindent Two conditions, two conditional branches (\texttt{cbnz}):

\begin{lstlisting}[language={}]
f:
    sub   sp, sp, #32
    stp   x29, x30, [sp, #16]
    add   x29, sp, #16
    stur  w0, [x29, #-4]
    str   w1, [sp, #8]
    ldur  w8, [x29, #-4]
    cbnz  w8, .LBB0_3
    b     .LBB0_1
.LBB0_1:
    ldr   w8, [sp, #8]
    cbnz  w8, .LBB0_3
    b     .LBB0_2
.LBB0_2:
    bl    act
    b     .LBB0_3
.LBB0_3:
    ldp   x29, x30, [sp, #16]
    add   sp, sp, #32
    ret
\end{lstlisting}

\subsection*{B.2 Example 1 at \texttt{-O3}, AArch64}

\noindent Both comparisons fused into \texttt{orr}, one conditional branch
(\texttt{cbz}):

\begin{lstlisting}[language={}]
f:
    orr   w8, w1, w0
    cbz   w8, .LBB0_2
    ret
.LBB0_2:
    b     act
\end{lstlisting}

\subsection*{B.3 Example 2 at \texttt{-O0}, AArch64}

\noindent A compare and a conditional branch around the assignment:

\begin{lstlisting}[language={}]
g:
    sub   sp, sp, #16
    str   w0, [sp, #12]
    str   w1, [sp, #8]
    ldr   w8, [sp, #12]
    cbnz  w8, .LBB0_2
    b     .LBB0_1
.LBB0_1:
    mov   w8, #42
    str   w8, [sp, #4]
    b     .LBB0_3
.LBB0_2:
    ldr   w8, [sp, #8]
    str   w8, [sp, #4]
    b     .LBB0_3
.LBB0_3:
    ldr   w8, [sp, #4]
    str   w8, [sp, #8]
    ldr   w0, [sp, #8]
    add   sp, sp, #16
    ret
\end{lstlisting}

\subsection*{B.4 Example 2 at \texttt{-O3}, AArch64}

\noindent No branch; a \texttt{csel} conditional-select instruction:

\begin{lstlisting}[language={}]
g:
    mov   w8, #42
    cmp   w0, #0
    csel  w0, w8, w1, eq
    ret
\end{lstlisting}

\subsection*{B.5 Example 2 at \texttt{-O3}, x86-64}

\noindent No branch; a \texttt{cmovne} conditional move:

\begin{lstlisting}[language={}]
g:
    testl    %edi, %edi
    movl     $42, %eax
    cmovnel  %esi, %eax
    retq
\end{lstlisting}

\subsection*{B.6 Example 3: macro provenance at \texttt{-g -O2}, AArch64}

\noindent Source (\texttt{ex3-macro-provenance.c}):
\begin{lstlisting}
void act(int);

#define CHECK(x, y)      \
    if ((x) > 0) {       \
        if ((y) > 0) {   \
            act(1);      \
        } else {         \
            act(2);      \
        }                \
    }

void h(int a, int b)
{
    CHECK(a, b);
}
\end{lstlisting}

\noindent Function body of the generated assembly with its line-table
directives (\texttt{.loc} \emph{file line column}): every instruction of the
first condition carries the invocation location 14:5, while the second
condition and the \texttt{cinc} formed from the inner branch carry line~0.
Unlike Listings B.1--B.5, this excerpt is deliberately reproduced verbatim
--- the \texttt{.loc} directives and \texttt{DEBUG\_VALUE} comments are
themselves the evidence under discussion. The complete file, including the
DWARF section boilerplate elided here, ships in the arXiv source package:
\begin{lstlisting}[language={},lastline=40,commentstyle=\color{black!90}]
	.file	"ex3-macro-provenance.c"
	.text
	.globl	h                               // -- Begin function h
	.p2align	2
	.type	h,@function
h:                                      // @h
.Lfunc_begin0:
	.file	0 "." "code\\ex3-macro-provenance.c" md5 0x14ef09637b754e23e0c20be4e6f54368
	.cfi_sections .debug_frame
	.cfi_startproc
// %bb.0:
	//DEBUG_VALUE: h:a <- $w0
	//DEBUG_VALUE: h:b <- $w1
	.file	1 "code" "ex3-macro-provenance.c" md5 0x14ef09637b754e23e0c20be4e6f54368
	.loc	1 14 5 prologue_end             // code/ex3-macro-provenance.c:14:5
	cmp	w0, #1
	b.lt	.LBB0_2
.Ltmp0:
// %bb.1:
	//DEBUG_VALUE: h:b <- $w1
	//DEBUG_VALUE: h:a <- $w0
	.loc	1 0 5 is_stmt 0                 // code/ex3-macro-provenance.c:0:5
	cmp	w1, #1
	mov	w8, #1                          // =0x1
	cinc	w0, w8, lt
.Ltmp1:
	//DEBUG_VALUE: h:a <- [DW_OP_LLVM_entry_value 1] $w0
	.loc	1 14 5 is_stmt 1                // code/ex3-macro-provenance.c:14:5
	b	act
.Ltmp2:
.LBB0_2:
	//DEBUG_VALUE: h:b <- $w1
	//DEBUG_VALUE: h:a <- $w0
	.loc	1 15 1                          // code/ex3-macro-provenance.c:15:1
	ret
.Ltmp3:
.Lfunc_end0:
	.size	h, .Lfunc_end0-h
	.cfi_endproc
                                        // -- End function
\end{lstlisting}

\noindent The same effect decoded from the assembled object file's
\texttt{.debug\_line} section (\texttt{llvm-objdump -d -l -r} on the
\texttt{-c} output of the identical command): the instructions of the second
condition carry no source-line annotation --- they lie in the line-0 region
--- while the surrounding instructions map to line~14 and~15; the relocation
line resolves the tail call:
\begin{lstlisting}[language={},commentstyle=\color{black!90}]

ex3.o:	file format elf64-littleaarch64

Disassembly of section .text:

0000000000000000 <h>:
; h():
; .\code\ex3-macro-provenance.c:14
       0:      	cmp	w0, #0x1
       4:      	b.lt	0x18 <h+0x18>
       8:      	cmp	w1, #0x1
       c:      	mov	w8, #0x1                // =1
      10:      	cinc	w0, w8, lt
; .\code\ex3-macro-provenance.c:14
      14:      	b	0x14 <h+0x14>
		0000000000000014:  R_AARCH64_JUMP26	act
; .\code\ex3-macro-provenance.c:15
      18:      	ret
\end{lstlisting}

\section*{Disclosure and trademarks}

The authors are employees of Accemic Technologies GmbH, which is the
applicant of WO~2026/162831~A1 and develops tooling based on the disclosed
approach; A.~Weiss is the named inventor of that application.

{\footnotesize CEDARtools and Accemic are trademarks of Accemic Technologies
GmbH. Arm and CoreSight are trademarks or registered trademarks of Arm
Limited (or its subsidiaries) in the US and/or elsewhere. Intel is a
trademark of Intel Corporation or its subsidiaries. RISC-V, RISC-V
International, and the RISC-V logos are trademarks of RISC-V International.
Infineon is a trademark of Infineon Technologies AG. All other product names,
company names and trademarks mentioned in this document are the property of
their respective owners and are used for identification purposes only; such
use does not imply affiliation with, sponsorship by, or endorsement from
their respective owners.\par}

\vspace{4pt}

\noindent The statements above about what a protocol transmits, what a debug
format promises and what a standard requires are based on the primary
documents cited below. The message-level MCDS documentation available to the
authors is subject to a non-disclosure agreement; this article therefore
names only what Infineon has published itself.

\end{document}